\documentstyle[aps,prb]{revtex}

\def \beq{\begin{equation}}
\def \eeq{\end{equation}}
\def \beqa{\begin{eqnarray}}
\def \eeqa{\end{eqnarray}}
\def \A{\mbox{\boldmath ${\cal A}$}}

\def \H{{\cal H}}
\def \a{{\rm \bf a}}
\def \e{\epsilon}

\def \Half{{\textstyle{1\over 2}}}
\def \bH{{\bf H}}
\def \bI{{\bf I}}
\def \bK{{\bf K}}

\def \bKt{{\bf \widetilde K}}
\def \x{{\bf x}}

\def \d{\delta}
\def \pd{\partial}
\def \tr{\tilde{\rho}}
\def \tz{\tilde{z}}

\begin{document}
\title{               \begin{flushleft}
{\footnotesize BU-CCS-941002}\\
{\footnotesize chem-ph/9411003}\\
{\footnotesize To Appear in {\it Computers in Physics} (1994)}
                      \end{flushleft}
Dimensional Perturbation Theory on the Connection Machine}
\author{Timothy C. Germann and Dudley R. Herschbach}
\address{Department of Chemistry, Harvard University,
     12 Oxford Street, Cambridge, MA 02138}
\author{Bruce M. Boghosian}
\address{Center for Computational Science, Boston University, 3
     Cummington Street, Boston, MA 02215}
\maketitle

\begin{abstract}
A recently developed linear algebraic method for the computation of
perturbation expansion coefficients to large order is applied to the
problem of a hydrogenic atom in a magnetic field.  We take as the zeroth
order approximation the $D \rightarrow \infty$ limit, where $D$ is the
number of spatial dimensions.  In this pseudoclassical limit, the
wavefunction is localized at the minimum of an effective potential
surface.  A perturbation expansion, corresponding to harmonic
oscillations about this minimum and higher order anharmonic correction
terms, is then developed in inverse powers of $(D-1)$ about this limit,
to 30th order.  To demonstrate the implicit parallelism of this method,
which is crucial if it is to be successfully applied to problems with
many degrees of freedom, we describe and analyze a particular
implementation on massively parallel Connection Machine systems (CM-2
and CM-5).  After presenting performance results, we conclude with a
discussion of the prospects for extending this method to larger systems.
\end{abstract}

\section{Introduction}
The spatial dimension has long been treated as a variable parameter in
analyzing critical phenomena and in other areas of physics.\cite{witten}
However, only in the past ten years has this concept been extensively
applied to atomic and molecular systems, particularly to develop
dimensional scaling methods for electronic structure.\cite{dscal} The
motivation for this unconventional approach is that the Schr\"odinger
equation reduces to easily solvable forms in the limits $D \rightarrow
1$ and/or $D \rightarrow \infty$.  When both limiting solutions are
available, interpolation in $1/D$ may be used to approximate the
physically meaningful $D=3$ result; this has yielded excellent results
for correlation energies of two-electron atoms\cite{loeser} and for
$H_2$ Hartree-Fock energies.\cite{tan} Alternatively, if the $D
\rightarrow \infty$ solutions are avaliable for both the problem of
interest and a simpler model problem (e.g., Hartree-Fock) for which $D =
3$ results are easier to calculate, the latter may be used to
``renormalize'' some parameter (e.g., nuclear charge). Then the $D
\rightarrow \infty$ solution with the renormalized parameter may give a
good approximation to the $D=3$ solution with the actual parameter
value.\cite{renorm}

Our work deals with another widely applicable dimensional scaling
method, a perturbation expansion in inverse powers of $D$ or a related
function, about the solution for the $D \rightarrow \infty$ limit.  That
limit is pseudoclassical and readily evaluated, as it reduces to the
simple problem of minimizing an effective potential
function.\cite{dscal} For large but finite $D$, the first-order
correction accounts for harmonic oscillations about this minimum, and
higher-order terms provide anharmonic corrections.  Dimensional
perturbation theory has been applied quite successfully to the
ground\cite{helium} and some excited states\cite{david93} two-electron
atoms and to the hydrogen molecule-ion\cite{h2plus} using a ``moment
method'' to solve the set of perturbation equations.  However, this
method is not easily extended to larger systems. It also requires a
different program for each eigenstate, and does not directly provide an
expansion for the wavefunction.

A recently developed linear algebraic method has overcome these
shortcomings.\cite{dunn93} This is conceptually quite simple, so can
easily be applied to systems with any number of degrees of freedom.  It
permits calculation of ground and excited energy levels using a single
program and the wavefunction expansion coefficients are directly
obtained in the course of computing the perturbation expansion for the
energy.  This method has thus far been applied to central force
problems, including quasibound states for which the complex eigenenergy
represents both the location and width of the resonance.\cite{tcg93}

The linear algebraic version of dimensional perturbation theory is also
well suited to parallel computation.  Here we demonstrate this for a
prototype problem with two degrees of freedom, the hydrogen atom in a
magnetic field.  This system has received much attention; it exhibits
chaotic behavior and poses difficulties that have challenged many
theoretical techniques.  Most theoretical approaches treat either the
magnetic field or the Coulomb potential as a perturbation and therefore
work best near either the low- or high-field limit, respectively.
However, the leading terms of a perturbation expansion in inverse powers
of $D$ include major portions of the nonseparable interactions in all
field strengths.  The efficacy of methods equivalent to the $1/D$
expansion has been demonstrated for the hydrogen atom in a magnetic
field\cite{bender82} and for kindred problems with an electric field or
crossed electric and magnetic fields,\cite{popov} although not in
formulations suited to parallel computation.  The present paper is
devoted solely to implementing of the linear algebraic method on the
Connection Machine, and to evaluating the performance of the
computational algorithm as well as prospects for treating systems with
more degrees of freedom.  Numerical results for the ground and several
excited states over a wide range of field strengths will be presented in
a separate paper.\cite{next}

\section{Theory}
The Schr\"odinger equation for a nonrelativistic hydrogenic atom in a
uniform magnetic field along the $z$ axis, in atomic units and
cylindrical coordinates, is given by
\beq
\left\{ -\frac{1}{2} \left[\frac{1}{\rho} \frac{\pd}{\pd \rho} \left(
  \rho \frac{\pd}{\pd \rho} \right) +
  \frac{\pd^2}{\pd z^2} - \frac{m^2}{\rho^2}\right] + \frac{mB}{2}
  + \frac{B^2 \rho^2}{8} - \frac{Z}{\sqrt{\rho^2 + z^2}}
\right\} \Psi(\rho,z) = E \Psi(\rho,z).
\eeq
Here $m$ is the azimuthal quantum number and the magnetic field $B$ is
measured in units $m_e^2 e^3 c/\hbar^3
\simeq 2.35 \times 10^9$ G.\cite{friedrich}  We can eliminate the
first derivative term with the substitution
\beq
\Phi(\rho,z) = \rho^{1/2} \Psi(\rho,z),
\eeq
which gives
\beq
\left\{ -\frac{1}{2}
  \left(
    \frac{\pd^2}{\pd \rho^2} +
    \frac{\pd^2}{\pd z^2}
  \right)
  + \frac{m^2-\frac{1}{4}}{2\rho^2} + \frac{mB}{2} + \frac{B^2 \rho^2}{8}
  - \frac{Z}{\sqrt{\rho^2 + z^2}}
\right\} \Phi(\rho,z) = E \Phi(\rho,z).
\label{eq:d3se}
\eeq
The two good quantum numbers are $m$ and the $z$ parity $\pi_z$,
although the observation of exponentially small level anticrossings
suggests a ``hidden'' approximate symmetry.\cite{herrick82} The trivial
(i.e. diagonal in each $m^\pi_z$ subspace) $mB/2$ term may be dropped,
to be added at the end of the calculation.

Now we consider the more general case where $\rho$ is the radius of a
$(D-1)$-dimensional hypersphere (with $D-2$ remaining angles), so that
the total number of spatial dimensions is $D$ when we include the
component $z$ parallel to the magnetic field.  In a manner completely
analogous to that described above, we may eliminate the first derivative
in the radial part of the Laplacian,
\[
\frac{1}{\rho^{D-2}} \frac{\pd}{\pd \rho} \left( \rho^{D-2}
  \frac{\pd}{\pd \rho} \right),
\]
by introducing the substitution
\beq
\Phi(\rho,z) = \rho^{(D-2)/2} \Psi(\rho,z),
\eeq
which leads to the same form as Eq.~(\ref{eq:d3se}), with $m^2 - 1/4$
replaced by $\Lambda(\Lambda+1)$ where $\Lambda = |m| +
\Half(D-4)$.\cite{herrick75} As before, we drop the $mB/2$ term.  Since
the resulting equation depends on $m$ and $D$ only through $\Lambda$, we
see that interdimensional degeneracies\cite{herrick75} connect states
$(|m|,D)$ with $(|m|-1,D+2)$.  It proves convenient to define a scaling
parameter as $\kappa \equiv D + 2|m| - 1$ and take $\Lambda =
\Half(\kappa-3)$.  The entire $|m| = 0,1,2,\ldots$ spectrum for the real
three-dimensional system can then be recovered simply by obtaining the
solutions for $\kappa = 2,4,6,\ldots$, respectively.

On introducing dimensionally scaled variables defined by
\beq
\rho = \kappa^2 \tr, \hspace{1cm}
z = \kappa^2 \tz, \hspace{1cm}
\tilde{B} = \kappa^3 B, \hspace{1cm}
\e = \kappa^2 E,
\label{eq:scal}
\eeq
we obtain the dimension-dependent Schr\"odinger equation
\beq
\left\{ -\frac{1}{2}  \d^2
  \left(
    \frac{\pd^2}{\pd \tr^2} +
    \frac{\pd^2}{\pd \tz^2}
  \right)
  + \frac{1-4\d + 3\d^2}{8\tr^2} + \frac{\tilde{B}^2 \tr^2}{8}
  - \frac{Z}{\sqrt{\tr^2 + \tz^2}}
\right\} \Phi(\tr,\tz) =
\e \Phi(\tr,\tz),
\label{eq:dimschro}
\eeq
where $\d = 1/\kappa$ is treated as a continuous perturbation
parameter.  The same form may be obtained simply by replacing
$|m|$ in Eq.~(\ref{eq:d3se}) by $(\kappa-2)/2$; this is the
procedure employed by Bender and coworkers.\cite{bender82}

We see that in the $\d \rightarrow 0$ limit, the problem reduces to
finding the minimum of the effective potential
\beq
V_{\mbox{eff}} (\tr,\tz) = \frac{1}{8\tr^2} + \frac{\tilde{B}^2 \tr^2}{8}
  - \frac{Z}{\sqrt{\tr^2 + \tz^2}}.
\eeq
Clearly, the effective potential is minimized for $\tz = z_m = 0$, so we
are left with the straightforward problem of minimizing a function of
one variable to determine $\rho_m$ and $V_{\mbox{eff}} (\rho_m, z_m)$.
In the zero-field limit, $\rho_m = (4Z)^{-1}$ and $V_{\mbox{eff}}
(\rho_m, z_m) = -2Z^2$ in the scaled energy units of
Eq.~(\ref{eq:scal}); in unscaled units this becomes
\beq
E = \frac{-Z^2}{2(|m|+1)^2},
\eeq
the correct expression for the ground state energy in each azimuthal
manifold. In the strong-field limit, $\rho_m \approx \tilde{B}^{-1/2}$
and $V_{\mbox{eff}} (\rho_m, z_m) \approx \tilde{B}/4$, or in unscaled
units
\beq
E \approx \frac{1}{2}B(|m|+1),
\eeq
which is simply the continuum threshold in each $m^{\pi_z}$ subspace.
This appropriate behavior in both limits, which has been noted
previously,\cite{bender82} is the motivation behind the definition of
$\kappa$ which we have used.\cite{shifted}

For large but finite $\kappa$ the system will undergo harmonic
oscillations about this minimum, so we introduce dimension-scaled
displacement coordinates $x_1$ and $x_2$ through
\beq
\tr = \rho_m + \d^{1/2}x_1, \hspace{1cm}
\tz = \d^{1/2}x_2.
\eeq
Substituting into Eq.~(\ref{eq:dimschro}) leads to
\beqa
 \left\{ -\frac{1}{2}  \d
  \left(
    \frac{\pd^2}{\pd x_1^2} +
    \frac{\pd^2}{\pd x_2^2}
  \right)
  + \d \left( \frac{1}{2} \omega_1^2 x_1^2
       + \frac{1}{2} \omega_2^2 x_2^2 + v_{0,0} \right)
  + V_{\mbox{eff}} (\rho_m, z_m) +
\right. && \nonumber \\   \left.
  \d \sum_{j=1}^\infty \d^{j/2} \left[ \left(
    \sum_{l=0}^{(j+2)/2} \; ^lv_{j,j+2} x_1^{j+2-2l} x_2^{2l} \right) +
    v_{j,j} x_1^j +
    v_{j,j-2} x_1^{j-2} \right]
\right\} && \Phi(x_1, x_2) =
\e \Phi(x_1, x_2),
\label{eq:normalse}
\eeqa
where
\beqa
\omega_1^2 &=& \frac{3}{4 \rho_m^4} - \frac{2 Z}{\rho_m^3} +
     \frac{\tilde{B}^2}{4},\nonumber\\
\omega_2^2 &=& \frac{Z}{\rho_m^3},\nonumber\\
v_{0,0} &=& - \frac{3}{4 \rho_m^2}, \nonumber\\
v_{1,-1} &=& 0.\nonumber
\eeqa
We next expand the wavefunction and energy in Eq.~(\ref{eq:normalse}) as
\beqa
\e &=& V_{\mbox{eff}} (\rho_m, z_m) + \d \sum_{j=0}^\infty
   \e_{2j} \d^j, \label{eq:e_exp}\\
\Phi(x_1, x_2) &=& \sum_{j=0}^\infty \Phi_j(x_1,x_2) \d^{j/2}
\eeqa
and collect powers of $\d^{1/2}$ to obtain an infinite set of
inhomogeneous differential equations,
\beq
\sum_{j=0}^p(\H_j-\e_j)\Phi_{p-j}=0,\quad p=0,1,2,\dots,
\label{eq:setofeq}
\eeq
where
\begin{mathletters}
\label{hams}
\beqa
\H_0&=&-\frac{1}{2}
  \left(
    \frac{\pd^2}{\pd x_1^2} +
    \frac{\pd^2}{\pd x_2^2}
  \right)
  + \frac{1}{2} \omega_1^2 x_1^2
       + \frac{1}{2} \omega_2^2 x_2^2 + v_{0,0} \\
\label{ham0}
\H_{j>0}&=&
    \left( \sum_{l=0}^{(j+2)/2} \; ^lv_{j,j+2} x_1^{j+2-2l} x_2^{2l} \right) +
    v_{j,j} x_1^j +
    v_{j,j-2} x_1^{j-2}
\label{hamj}
\eeqa
\end{mathletters}\noindent
with $\e_{2i+1}=0$.  The case $p=0$ corresponds to a pair of independent
harmonic oscillators, with the solution
\beqa
&&\e_0=\left(\nu_1+\Half\right)\omega_1+\left(\nu_2+\Half\right)\omega_2+
   v_{0,0}, \\
&&\Phi_0(x_1,x_2)=h_{\nu_1}(\omega_1^{1/2} x_1)h_{\nu_2}(\omega_2^{1/2}
x_2),
\eeqa
where $\nu_1$ and $\nu_2$ are quantum numbers for the normal modes
corresponding to motion perpendicular to and parallel to the magnetic
field, respectively, and the $h_i$ are the one-dimensional harmonic
oscillator eigenfunctions.  The correspondences between these
large-dimension quantum numbers and the conventional labels for the low,
intermediate, and high field cases will be detailed
elsewhere.\cite{next}

\section{Tensor method}
The higher-order terms in the energy and wavefunction expansions may be
computed using the linear algebraic method.\cite{dunn93} The
wavefunction expansion terms $\Phi_j(x_1,x_2)$ are expanded in terms of
the harmonic oscillator eigenfunctions $h_i$,
\beq
\Phi_j(x_1,x_2) = \sum_{i_1} \sum_{i_2} \; ^{i_1,i_2}a_j
   h_{i_1}(\omega_1^{1/2} x_1)h_{i_2}(\omega_2^{1/2} x_2),
\eeq
so that the representation $\a_j$ is a tensor of rank 2.  The
displacement coordinates $x_i$ are represented in this basis by the
matrix
\beq
\x_i = \frac{1}{\sqrt{2\omega_i}} \left( \begin{array}{ccccc}
0 & \sqrt{1} & 0 & 0 & \\
\sqrt{1} & 0 & \sqrt{2} & 0 & \\
0 & \sqrt{2} & 0 & \sqrt{3} & \cdots \\
0 & 0 & \sqrt{3} & 0 &  \\
 & & \vdots &  & \ddots
\end{array} \right),
\label{eq:xrep}
\eeq
and the Hamiltonian $\H_{j>0}$ is represented as a linear combination of
direct (outer) products of these matrices, namely
\beq
\bH_{j>0}= \left(
    \sum_{l=0}^{(j+2)/2} \; ^lv_{j,j+2} \x_1^{j+2-2l} \otimes \x_2^{2l}
    \right) + v_{j,j} \x_1^j \otimes \bI +
    v_{j,j-2} \x_1^{j-2} \otimes \bI.
\eeq

Using these representations, the $j=0$ term of Eq.~(\ref{eq:setofeq}),
$(\H_0 - \epsilon_0)\Phi_p$, becomes
\beq
(\bH_0 - \epsilon_0 \bI)\a_p = ((\omega_1 \bK_1) \oplus (\omega_2 \bK_2))
\a_p,
\eeq
where $\bK_i$ is a diagonal matrix with element $(j,j)$ equal to
$j-\nu_i$, since the basis functions $h_{i_1}(\omega_1^{1/2}
x_1)h_{i_2}(\omega_2^{1/2} x_2)$ are eigenfunctions of $\H_0$.  Let us
pause at this point to be sure that there is no confusion regarding the
notation.  A ``direct sum'' is analogous to the direct product
operation, i.e. ${\bf C} = {\bf A} \oplus {\bf B}$, where ${\bf A}$ and
${\bf B}$ are matrices, gives a rank 4 tensor with elements $C_{j_1 k_1
j_2 k_2} = A_{j_1 k_1} + B_{j_2 k_2}$.  Then the multiplication ${\bf E}
= {\bf C} {\bf D}$, where ${\bf D}$ and ${\bf E}$ are rank 2 tensors,
implies $E_{j_1 j_2} = C_{j_1 k_1 j_2 k_2} D_{k_1 k_2}$, where we are
using the implied summation convention.  If we define a rank 4 tensor
$\bKt$ with elements
\beq
\bKt_{j_1 k_1 j_2 k_2} = \left\{
\begin{array}{ll}
1, & j_1 = k_1 = \nu_1, j_2 = k_2 = \nu_2 \\
(\omega_1(j_1 - \nu_1) + \omega_2 (j_2 - \nu_2))^{-1}, &
j_1 = k_1 , j_2 = k_2 ,
j_1 \neq \nu_1 \;\; \mbox{and/or} \;\; j_2 \neq \nu_2 \\
0, & \mbox{otherwise}
\end{array} \right.,
\label{eq:Ktilde}
\eeq
then it may be verified that $\bKt((\omega_1 \bK_1) \oplus (\omega_2
\bK_2))$ is equal to the direct product $\bI \otimes \bI$ except for a
zero in element $\nu_1 \nu_1 \nu_2 \nu_2$.  However, due to the
orthogonality condition
\beq
\a_0^T \a_p = \delta_{0,p},
\label{eq:ortho}
\eeq
where $\delta_{0,p}$ is the Kronecker delta, we have
\beq
\bKt  ((\omega_1 \bK_1) \oplus (\omega_2 \bK_2)) \a_p = \a_p,
\hspace{0.5in} p > 0.
\eeq
Therefore if we multiply Eq.~(\ref{eq:setofeq}) on the left by $\bKt$, we
obtain a recursive solution for $\a_p$,
\beq
\a_p = - \bKt \sum_{j=1}^p (\bH_j - \epsilon_j \bI) \a_{p-j}.
\label{eq:ap}
\eeq
If we instead multiply Eq.~(\ref{eq:setofeq}) on the left by $\a_0^T$
and use the orthogonality condition of Eq.~(\ref{eq:ortho}), we obtain
an expression for $\epsilon_p$ in terms of $\a_{j<p}$,
\beq
\epsilon_p = \sum_{j=1}^p \a_0^T \bH_j \a_{p-j}.
\label{eq:ep}
\eeq

{}From a computational standpoint, the only operation with which we need
to concern ourselves is the multiplication of a rank 4 tensor with a
rank 2 tensor, $\bH_j \a_{p-j}$.  According to Eq.~(\ref{eq:Ktilde}),
the nonzero elements of $\bKt$ appear in Eq.~(\ref{eq:ap}) simply as
scaling factors for each element of $\a_p$ after the summation has been
completed.

\section{Implementation}
\subsection{Rewriting the recursion relations}
In order to solve Eqs.~(\ref{eq:ap}) and (\ref{eq:ep}) recursively, we
introduce the rank 2 tensor $\A_{jln}$ which is defined as
\beq
\A_{jln} \equiv (\x_1^{j-2l} \otimes \x_2^{2l}) \a_n,
\label{eq:def}
\eeq
so that the $\bH_j \a_{p-j}$ term which most concerns us may be
expressed as a linear combination of these tensors, namely
\beq
\bH_j \a_{p-j} = \left(
    \sum_{l=0}^{(j+2)/2} \; ^lv_{j,j+2} \A_{j+2,l,p-j}
    \right) + v_{j,j} \A_{j,0,p-j} +
    v_{j,j-2} \A_{j-2,0,p-j}.
\label{eq:H_a}
\eeq
In order for this change of notation to be beneficial, we need to find
some simple recursion relation(s) relating the $\A_{jln}$ tensors for a
given order to those tensors which have been used at lower orders.  In
addition, we hope that only a small subset of the tensors used for
previous orders are needed, so that a managable number of these tensors
(and the wavefunction tensors $\a_n$) need to be allocated in memory.
Two recursion relations are immediately evident, namely
\begin{mathletters}
\label{recurs}
\beqa
\A_{j+1,l,n} &=& (\x_1 \otimes \bI) \A_{jln},\label{recur1} \\
\A_{j+2,l+1,n} &=& (\bI \otimes \x_2^2) \A_{jln}. \label{recur2}
\eeqa
\end{mathletters}
It remains to be seen how many terms may be computed using these
relations, and conversely how many will need to be computed ``from
scratch.''  We show in Fig.~\ref{fig:matrices} the tensors $\A_{jln}$
which appear in the first term of Eq.~(\ref{eq:H_a}) for the first few
orders $p$, and also indicate which recursion relations (if any) may be
used to compute each tensor (with preference given to Eq.~(\ref{recur1})
if both are applicable).  We see that Eq.~(\ref{recur1}) simply
``shifts'' the staircase-like pattern of tensors to the right, while
Eq.~(\ref{recur2}) fills in all but three of the gaps left by this
shift.  Turning our attention to the second term of Eq.~(\ref{eq:H_a}),
we note that at order $p$ this term represents tensors which have
appeared in the first term at order $p-2$, with the exception of
$\A_{1,0,p-1}$ and $\A_{2,0,p-2}$.  Similarly, the third term of
Eq.~(\ref{eq:H_a}) at order $p$ involves tensors which occured in the
second term at order $p-2$, with the exception of $\A_{0,0,p-2}$, which
is simply the wavefunction tensor $\a_{p-2}$ which is resident in
memory.  Therefore, at each order we have at most five tensors which
cannot be computed from direct application of Eqs.~(\ref{recurs}).

In reality, the computation of these five tensors is no more difficult
than the two standard recursion relations:
\beqa
\A_{1,0,p-1} &=& (\x_1 \otimes \bI) \a_{p-1}, \nonumber \\
\A_{3,0,p-1} &=& (\x_1^2 \otimes \bI) \A_{1,0,p-1}, \nonumber\\
\A_{3,1,p-1} &=& (\bI \otimes \x_2^2) \A_{1,0,p-1}, \label{altrec}\\
\A_{2,0,p-2} &=& (\x_1^2 \otimes \bI) \a_{p-2}, \nonumber\\
\A_{4,2,p-2} &=& (\bI \otimes \x_2^2) [(\bI \otimes \x_2^2)\a_{p-2}].\nonumber
\eeqa
We note that these relations, as well as the two standard recursion
relations, only involve computation along one coordinate axis.

By properly ordering the computation steps, we may compute all of the
tensors $\A_{jln}$ at order $p$ ``in place'' (i.e. overwriting all of
the tensors from order $p-1$), thus minimizing the need for temporary
storage.  The second recursion relation uses tensors from order $p-2$,
which must be temporarily stored during the order $p-1$ calculation.
However, this relation is only used for $(p-2)/2$ tensors at order $p$,
so the additional storage cost is minor.  The algorithm may be
summarized as follows (see also Fig.~\ref{fig:algorithm}): (1) Recursion
Eq.~(\ref{recur1}) is applied to all tensors, working from right to left
in Fig.~\ref{fig:algorithm} to avoid the need for temporary storage.
Note that this step does {\em not} overwrite those tensors which we will
need later for Eq.~(\ref{recur2}).  (2) Recursion Eq.~(\ref{recur2}) is
applied to tensors from order $p-2$ which are stored separately.  By
again working from right to left, we may replace the tensors in
temporary storage as they are used with the corresponding tensors from
order $p-1$, so that both the main set of tensors and the temporary set
are updated as they are used.  (3) Five original tensors
(Eq.~(\ref{altrec})) are computed.  (4) $\a_p$ is computed by taking a
linear combination of tensors, and the contributions of those tensors
which are necessary for $\a_{p+2}$ and $\a_{p+4}$ are now computed.

\subsection{Implementing the recursion relations on the Connection Machine}
We have implemented the computation of the energy eigenvalues, as
described in the previous section, using a data-parallel algorithm on
the CM-5 Connection Machine computer.  The CM-5 computer consists of up
to 16,384 processor nodes, each consisting of a Sun Sparc processor and
four vector units for fast floating-point computation.  It supports both
the data-parallel and the message-passing styles of parallel
computation.

The data-parallel capabilities of the CM-5 are supported in the form of
high-level languages: C* is a data-parallel extension of the C
programming language; while CM Fortran is similar to the Fortran 90
draft standard, augmented by some features from High-Performance Fortran
(HPF) such as the {\tt FORALL} statement.  For this work, we chose to
use CM Fortran.

The basic parallel data type in CM Fortran is the usual Fortran array.
When the CM Fortran compiler encounters a {\tt DIMENSION} statement, it
allocates memory to spread the array across the processors of the CM-5.
If it is desired that certain axes of a multidimensional array be
localized to a single processor, that can be arranged by a simple
compiler directive, called the {\tt LAYOUT} directive.  Axes localized
in this way are called {\it serial axes}, while those spread across the
processor array are called {\it news axes}.

Operations on corresponding elements of arrays with identical layouts
can then be performed by each processor in parallel.  If there are more
array elements than physical processors, the compiler arranges for each
physical processor to contain a {\it subgrid} of multiple array
elements, and array operations are then multiplexed over these elements
as required.  The compiler's orchestration of this process is completely
transparent to the CM Fortran user, who may then regard each array
element as living in its own {\it virtual processor}.

In this picture, operations involving noncorresponding array elements
require interprocessor communication.  Data-parallel programming
languages, such as CM Fortran, support many different varieties of this.
For the problem at hand, however, only {\it nearest-neighbor
communication} is used.

Nearest-neighbor communication is needed when, for example, you would
like to take a linear combination of the array elements in (virtual)
processors, $i-1$, $i$, and $i+1$.  It is supported by a Fortran 90 (and
CM Fortran) intrinsic function known as {\tt CSHIFT}.  If {\tt A} is a
CM Fortran array, and {\tt i} and {\tt j} are integers, then {\tt
CSHIFT(A,i,j)} is another such array, whose elements are shifted along
axis {\tt i} by an amount {\tt j}.  There is another variant of this
function, known as {\tt EOSHIFT}, that does basically the same thing,
differing only in its treatment of the boundary elements of the array;
whereas {\tt CSHIFT} treats the boundaries as though the array were
periodic, {\tt EOSHIFT} has extra arguments for arrays of codimension
one to be inserted at the boundary.

Our storage method uses the fact that the position matrices in the
harmonic oscillator representation, given by Eq.~(\ref{eq:xrep}), are
doubly banded.  We store only the nonzero elements of the $i$th row in
(virtual) processor $i$.  Thus, one can think of the band below the
diagonal as a one-dimensional array, {\tt XL}, and that above the
diagonal as another one-dimensional array, {\tt XU}.  Then, {\tt XL(i)}
and {\tt XU(i)} contain the two nonzero elements in row {\tt i}, and we
can write
\[
{\bf x}_{\alpha\beta} = {\tt XL}(\alpha)\delta_{\alpha-1,\beta} +
                        {\tt XU}(\alpha)\delta_{\alpha+1,\beta}.
\]

The main operation needed for the implementation of the recursion
relations, Eqs.~(\ref{recurs}), is the inner product of the position
matrices with the matrices, $\A_{j,l,n}$.  The latter are stored as
dense three-dimensional arrays, with two {\it news} axes for the
components of the matrices, and one {\it serial} axis representing the
allowed combinations of $j$, $l$, and $n$.  Thus, one can imagine these
matrices as spread across the processor array, with corresponding
components localized to the same (virtual) processor.

Consider one of these matrices; call it $\A_{\mu\nu}$ (where the Greek
indices now label the {\it tensor} components and {\it not} the serial
axis, $\{j,l,n\}$, which is being supressed for the moment).  Then
Eq.~(\ref{recur1}), which is the inner product of ${\bf x}$ with (the
first component of) $\A$, is
\begin{equation}
(\x_{\xi\mu} \otimes \bI)\A_{\mu\nu} =
   {\tt XL}(\xi)\A_{\xi-1,\nu} +
   {\tt XU}(\xi)\A_{\xi+1,\nu}.
\label{eq:ip}
\end{equation}
Note that we can implement this in terms of two {\tt CSHIFT} operations
along the first axis of $\A$.  Since this must be done for all values of
$\mu$, we add this {\it instance axis} to the arrays, {\tt XL} and {\tt
XU}.

The complete algorithm for implementing the inner product then begins by
dimensioning the arrays as follows:
\begin{center}
\begin{verbatim}
      INTEGER m,n
      DIMENSION XL(nmax,nmax), XU(nmax,nmax), AA(nmax,nmax)
\end{verbatim}
\end{center}
The {\tt XL} and {\tt XU} arrays are then initialized as follows:
\begin{center}
\begin{verbatim}
      FORALL(m=1:nmax,n=1:nmax) XL(m,n) = SQRT(n-1)
      FORALL(m=1:nmax,n=1:nmax) XU(m,n) = SQRT(n)
\end{verbatim}
\end{center}
where the {\tt FORALL} statements are parallel {\tt DO}-loop
constructions supported in High-Performance Fortran (and CM Fortran).
For example, the first such statements cause each (virtual) processor to
take its own index, decrement it by one, and take the square root.
Alternatively, rather than allocate a separate array for {\tt XU}, we
can obtain it from {\tt XL} by a simple {\tt EOSHIFT} operation.

The inner product in Eq.~(\ref{eq:ip}) is then taken as follows:
\begin{center}
\begin{verbatim}
      XA = XL*EOSHIFT(AA,1,-1) + XU*EOSHIFT(AA,1,+1)
\end{verbatim}
\end{center}
Using this method, the recursion relations, Eqs.~(\ref{recurs}) and
(\ref{altrec}), can be implemented, and the terms in Eq.~(\ref{eq:H_a})
can be multiplied by the scalars, $v_{i,j}$, and summed in place.
(Since the $\{j,l,n\}$ axis is serial, the arrays $\A_{j,l,n}$ are all
aligned in the processor array.)  Similarly, the sum over $j$ in
Eq.~(\ref{eq:ep}) can be taken (without the premultiplication by ${\bf
a}_0$, which is independent of $j$.)

Finally, the result for $\epsilon_p$, given by Eq.~(\ref{eq:ep}), can be
found quite easily.  Since ${\bf a}_0$ is an eigenvector in our
representation, we can effectively premultiply it by simply taking the
first component of $\sum_j {\bf H}_j {\bf a}_{p-j}$.  The statement
\begin{center}
\begin{verbatim}
      ep = XA(1,1)
\end{verbatim}
\end{center}
reaches into the (virtual) processor containing the {\tt (1,1)}
component of the array {\tt XA}, pulls out its value, and writes it into
the scalar (stored on the control processor) quantity {\tt ep}, from
where it can be manipulated, output, etc.

Thus, the array extensions of Fortran 90 and High-Performance Fortran,
as embodied in the CM Fortran language, give a convenient set of
primitives for the efficient implementation of the dimensional
perturbation theory algorithm.

\subsection{Contraction of axis lengths}
All of the wavefunction tensors $\a_n$ and the tensors $\A_{jln}$ used
in their computation are allocated to the size of the final wavefunction
tensor, $\a_{p_{max}}$, which permits calculation of energy expansion
coefficients through $\epsilon_{p_{max}+1}$.  However, we have not yet
attempted to take advantage of the sparseness of these tensors.  To
illustrate the amount of room for improvement, Fig.~\ref{fig:sparse}
indicates the nonzero elements of the first few wavefunction tensors for
the case of a ground state ($\nu_1 = \nu_2 = 0$) calculation.  Since we
would lose much of the efficiency of the routines for the recursion
relations described above if we attempted to treat these as general
sparse tensors, we will instead look for smaller-scale improvements.

First, we see that since $\x_2$ only occurs in even powers, alternating
columns are always zero and need not be stored.  In addition to cutting
the storage in half, this reduces the two next-nearest neighbor
communications required for the $(\bI \otimes \x_2^2)$ multiplication to
two nearest neighbor communications.

Looking along the other axis, we see that either even or odd rows, but
not both in a given tensor, contain nonzero elements, so that the rows
may be paired up as long as we maintain some sort of flag telling us
whether the even or odd row of the pair is represented.  This also
carries an extra communication benefit in addition to the storage
improvement; one of the two nearest-neighbor communications in the
$(\x_1 \otimes \bI)$ multiplication is converted to an entirely local
operation.

\section{Performance}
We now turn to the mapping of these arrays to the individual processors
on the Connection Machine, in the typical case where the array size
exceeds the number of processors.  The recursion relation of
Eq.~(\ref{recur1}) may be applied independently to different columns
since the communication is entirely within columns.  Similarly,
Eq.~(\ref{recur2}) may operate independently on separate rows.  Since
the former recursion relation is applied much more often than the
latter, we would expect that the optimal array layout would be that in
which each processor operates on a small number of long column segments,
i.e. the first (row) axis is ``more serial'' while the second (column)
axis is ``more parallel.''

We can in fact obtain a quantitative measure of this balance.  Suppose
we have an $N \times N$ matrix to be allocated on a system of $M$
processors.  If we assign $n$ columns (or segments of columns) to each
processor, then the length of each of these column segments is $N^2/nM$,
as shown in Fig.~\ref{fig:layout}.  The first recursion relation may be
implemented in CM Fortran as described in section IV.\ B, with the
addition of an {\tt IF} clause due to
the contraction of the first axis, as described in the previous section.
Using appropriately defined arrays {\tt X0} and {\tt XLU}, this may be
accomplished as follows:
\begin{center}
\begin{verbatim}
      IF (odd rows nonzero) THEN
         XA = X0*AA + XLU*EOSHIFT(AA,1,-1)
      ELSE
         XA = X0*AA + EOSHIFT(XLU*AA,1,+1)
      ENDIF
\end{verbatim}
\end{center}
Whichever branch of this clause is taken, there are three arithmetic
operations on each array element, with a total cost of $3 N^2 t_A / M$,
where $t_A$ represents the time for a single arithmetic operation.  The
{\tt EOSHIFT} operation will move $n$ elements from any given processor
into an adjacent processor, while the remaining $N^2/M - n$ elements
require an on-processor move (except for the case $n = N/M$, where all
elements are moved on-processor).  Thus the {\tt EOSHIFT} time is given
by
\[
\begin{array}{ll}
n t_Q & \mbox{if $n = N/M$} \\
n t_Q + \left( \frac{N^2}{M} - n \right) t_M & \mbox{otherwise}
\end{array} ,
\]
where $t_Q$ represents the queue waiting time for interprocessor
communication and $t_M$ is the on-processor move time.

The second recursion relation can be effected by the CM Fortran
statement
\begin{center}
\begin{verbatim}
     XA = X20*AA + EOSHIFT(X2LU*AA,2,-1) + X2LU*EOSHIFT(AA,2,+1)
\end{verbatim}
\end{center}
where the definition of the constant arrays {\tt X20} and {\tt X2LU} is
dependent on the contraction of the second axis, as described in the
previous section.  The arithmetic cost is $5 N^2M t_A / M$ and the cost
of the two {\tt EOSHIFT} operations is
\[
\begin{array}{ll}
\frac{2 N^2}{nM} t_Q & \mbox{if $n = N$} \\
\frac{2 N^2}{nM} t_Q + 2 \left( \frac{N^2}{M} - \frac{N^2}{nM}
\right) t_M & \mbox{otherwise}
\end{array} .
\]

For a program which applies the first recursion relation $n_1$ times and
the second relation $n_2$ times, the total time incurred by the two
relations is
\begin{equation}
t_{recur} = \frac{N^2}{M} [ (3 n_1 + 5 n_2) t_A +
   (n_1 + 2 n_2) t_M ] + \left(n_1 n + \frac{2 n_2 N^2}{nM} \right)
   (t_Q - t_M),
\end{equation}
where we have neglected the two special cases for $n$ described above.
The first term is independent of the specific layout, described by the
parameter $n$.  Since the only interprocessor data motion in the entire
program arises from the recursion relations, the second term can be used
to determine the optimal array layout. Since $t_Q > t_M$, we want to
minimize the term $n_1 n + 2 n_2 N^2 / n M$.  Setting its derivative
with respect to $n$ equal to zero gives a simple expression for the
optimal layout parameter $n_{opt}$,
\begin{equation}
n_{opt} = \sqrt{\frac{2 n_2 N^2}{n_1 M}}.
\label{eq:nopt}
\end{equation}
For example, a 30th order calculation using the implementation discussed
here requires $N = 128$, $n_1 = 18441$, $n_2 = 925$.  A 64-processor
CM-5 contains 256 vector units (hence $M=256$).  Thus, we would estimate
$n_{opt} \approx 2.5$, in agreement with the empirical observation that
$n=2$ results in faster run times than either $n=1$ or $n=4$ on a
64-processor CM-5 (see Table I).  Special care must be paid to the
special case $n=N/M$ when $M \leq N$, for a layout with the first axis
entirely serial $(n=1)$ may be optimal despite the fact that
Eq.~(\ref{eq:nopt}) gives a larger value for $n_{opt}$.

Table I presents timings and the corresponding floating point rates for
20th and 30th order calculations.  The recursion relations involve
substantial interprocessor communication; for the problem sizes
considered here, we have found that approximately 40\% of the CM
execution time is spent performing arithmetic, the remainder being
consumed by communication operations.

In spite of this, we were able to obtain a performance of nearly 20
Mflops per CM-5 processor node for the largest subgrid sizes considered
here.\cite{note} Since the algorithm described here is fully scalable,
we expect that if the subgrid size was kept fixed and the problem was
ported to the largest CM-5's existing today (1024 processor nodes), the
total performance would be about 20 Gflops.

Keeping the subgrid size fixed, however, implies that the problem size
grows with the hardware.  Again referring to Table I, we see that if we
move a fixed-size problem to larger machines, the per-processor
performance degrades.  This is because the correspondingly decreasing
subgrid size means that a larger fraction of time is being spent on
latencies (overhead costs).

Thus, massively parallel implementations of this algorithm will be
useful if there is something to be gained by going to larger problem
sizes.  The most obvious way to do this is to increase the order of the
perturbation series, and hence the sizes of the arrays involved, the
number of recursion relations that must be used, and the accuracy of the
results.  Here, however, we run into a problem.  Table II shows that at
30th order we have exhausted (or nearly exhausted) the significant
digits of the IEEE standard double-precision floating-point numbers used
in the calculation.  Thus, to grow the problem in this direction, we
would need to employ quadruple-precision arithmetic.

There are, however, other options for growing the problem size in a
useful manner.  Table III shows results for several different magnetic
field strengths.  Indeed, one of the more interesting things to study in
chaotic systems of this sort is the continuous dependence of the energy
levels on the magnetic field strength, or other system parameter.
Variation of these parameters provides another dimension over which to
parallelize, and thereby maintain large subgrid sizes.

\section{Conclusions}
In Table II we give ground state expansion coefficients $\epsilon_{2j}$
for two field strengths, $B = 1$ and 1000, computed in double-precision
arithmetic.  Comparison with quadruple-precision calculations shows that
the accumulation of roundoff error causes a linearly decreasing number
of significant digits in the coefficients, most pronounced for those
series with the slowest rate of growth in the coefficients.

In their work, Bender, Mlodinow, and Papanicolaou applied Shanks
extrapolation to accelerate convergence.\cite{bender82} For the
higher-order series we have computed, the large order divergent behavior
due to singularities renders this method ineffective.\cite{b_and_o}
Consequently, we employ Pad\'e approximants to sum our series.  Table
III compares our $1/\kappa$ results for the ground state of the $m$ = 0
and $-1$ mainfolds with variational calculations\cite{rosner} and lower
and upper bounds,\cite{handy} where available.  As noted by Goodson and
Watson, the order at which roundoff error leaves no significant digits
in the expansion coefficient $\epsilon_{2j}$ may be determined by noting
qualitative changes in the root and ratio tests.\cite{david93} We also
give in Table III these maximum orders which are attainable using
double-precision arithmetic.  Results for excited states will be
presented elsewhere.\cite{next}

For general problems with $\tau$ degrees of freedom, the $\a_n$ and
$\A_n$ objects are tensors of rank $\tau$.  It is clear that storing all
of the tensor elements will rapidly become impractical; while a $128
\times 128$ array of double precision numbers consumes 131 kBytes of
memory, a $128 \times 128 \times 128 \times 128$ object requires 2 GB.
Even using a parallel I/O device such as the DataVault or Scalable Disk
Array for temporary storage, it is evident that it will become necessary
to incorporate the sparse nature of these data structures more
explicitly and to reduce the communication (especially to external
devices) at the expense of additional arithmetic wherever possible.

It has recently been shown\cite{dunn93} that the recursion relations may
be reordered so that only the $\a_n$ tensors need to be stored.  This
approach has been used in the computation of expansion coefficients for
the helium atom, a system with three degrees of freedom.  Numerical
estimates of the storage and computational costs associated with this
approach indicate that large order calculations (20th order or higher)
are at present restricted to at most six degrees of freedom, {\em e.g.}
a three-electron atom.\cite{dunn93} However, a promising approach may be
to combine low-order perturbation expansions with other methods.  In
particular, the recent development of expansions about the $D
\rightarrow 0$ limit\cite{bender92} may provide means to augment the $D
\rightarrow \infty$ results.  Expansions about both of these limits can
be enhanced by renormalization schemes.

\vspace{0.5in}
\begin{center} {\large \bf Acknowledgements} \end{center}

T.C.G. gratefully acknowledges the award of a Computational Science
Graduate Fellowship from the U.S. Department of Energy, and would like
to thank Thinking Machines Corporation for their hospitality during the
time this work was performed.

\clearpage
\begin{table}
\tighten
\begin{tabular}{lrrrr}
Architecture & $P$ & $n$ & CM time (sec) &
  MFLOPS/sec/node \\ \hline
\multicolumn{5}{c}{\sc 20th order calculation (N=64)} \\
CM-5 (VU) & 32 & 2 & 0.896 & 5.012 \\
          & 128 & 2 & 0.730 & 1.536 \\
\multicolumn{5}{c}{\sc 30th order calculation (N=128)} \\
CM-2 & 256 & 1 & 11.91  & 0.588 \\
     & 512 & 1 &  7.11  & 0.492 \\
CM-5 (VU) & 4 & 1 & 22.78  & 19.668 \\
     & 32 & 1 &  4.62 & 12.124 \\
     & 64 & 1 &  3.81 \\
     & 64 & 2 &  3.68 & 7.608 \\
     & 64 & 4 &  3.75 \\
     & 128 & 2 &  2.82 & 4.964 \\
\end{tabular}
\caption{Performance results on Connection Machines with $P$ nodes,
where the optimal array layout is described by $n$ (see text).  For the
CM-2 using the slicewise execution model, each floating-point processor
node is comprised of 32 bit-serial processors and a floating-point
accelerator chip, so a full 64K CM-2 has 2048 processor nodes ($M=P$).
For the CM-5, each processing node (PN) has four vector units, so $M =
4P$ in predicting performance.}
\end{table}

\clearpage
\begin{table}
\tighten
\begin{tabular}{rdd}
$j$ & \multicolumn{1}{r}{\makebox[1in][l]{$\epsilon_{2j} \;(B=1)$}} &
\multicolumn{1}{r}{\makebox[1in][l]{$\epsilon_{2j} \;(B=1000)$}} \\ \hline
$-$1 &  $-$1.577218587578393	&	1.910051627706109(3)\\
0 &     0.63293278553615$_1$	&       3.61765949346725$_3$(2)\\
1 &  $-$0.3281655376303$_1$	&    $-$1.61755795476809$_6$(3)\\
2 &     0.189180754067$_1$	&       8.71028965022799$_1$(3)\\
3 &  $-$0.1202775104$_2$	&    $-$5.8564694134795$_7$(4)\\
4 &     0.10221091$_4$		&       4.518169171696$_0$(5)\\
5 &  $-$0.1685588$_6$		&    $-$2.55253312427$_6$(6)\\
6 &     0.46534$_7$		&    $-$4.75598614937$_9$(7)\\
7 &  $-$1.486$_3$		&       3.36811883775$_9$(9)\\
8 &     4.92$_6$		&    $-$1.4285156162$_5$(11)\\
9 &  $-$1.6$_8$(1)		&       5.3728029857$_7$(12)\\
10 &   1.$_0$(2)		&    $-$1.876673877$_8$(14)\\
11 & &       5.834807058$_2$(15)\\
12 & &    $-$1.29765909$_0$(17)\\
13 & &    $-$1.39749835$_0$(18)\\
14 & &       4.7606391$_8$(20)\\
15 & &    $-$4.5262663$_0$(22)\\
16 & &       3.365746$_4$(24)\\
17 & &    $-$2.179148$_6$(26)\\
18 & &       1.21980$_8$(28)\\
19 & &    $-$5.17984$_0$(29)\\
20 & &       3.7664$_0$(30)\\
21 & &       2.6805$_1$(33)\\
22 & &    $-$4.395$_8$(35)\\
23 & &       5.073$_1$(37)\\
24 & &    $-$4.92$_4$(39)\\
25 & &       4.09$_7$(41)\\
26 & &    $-$2.6$_8$(43)\\
27 & &       7.8$_6$(44)\\
28 & &       1.$_5$(47)\\
29 & &    $-$4.$_0$(49)\\
\end{tabular}
\caption{Expansion coefficients $\epsilon_{2j}$ for the ground-state energy,
defined in Eq.(\protect\ref{eq:e_exp}), for field
strengths $B=1$ and 1000. Note that due to
the scaling of Eq.(\protect\ref{eq:scal}), $\epsilon$
should be divided by $(D-1)^2 = 4$ to give the $m=0$ energy $E$. Subscripts
specify the digit after the last digit which is expected to be
significant, and the number in parenthesis following each entry is the
power of 10 multiplying that entry.}
\end{table}

\clearpage
\begin{table}
\tighten
\begin{tabular}{rrdddcl}
\multicolumn{1}{c}{$B$} &
\multicolumn{1}{c}{order} &
\multicolumn{1}{c}{Present work} &
\multicolumn{1}{c}{R\"osner {\em et al.}\protect\cite{rosner}} &
\multicolumn{3}{c}{Handy {\em et al.}\protect\cite{handy}} \\ \hline
\multicolumn{6}{c}{$E_B, \;1s_0$ state}\\
0.1 & 7 & 0.54752648 & 0.547527 & & & \\
1   & 11 & 0.831169 & 0.831169 & & &  \\
2   & 12 & 1.022213 & 1.022214 & 1.0222138 &$<E_B<$& 1.0222142  \\
20 & 20 & 2.21539 & 2.215399 & 2.215325 &$<E_B<$& 2.215450 \\
200 & 24 & 4.7275&  4.727 & 4.710 &$<E_B<$& 4.740\\
300 & 25 & 5.362 & 5.361 & 5.34 &$<E_B<$& 5.39 \\
1000 & 30 & 7.67 & 7.662 & 7.55 &$<E_B<$& 7.85 \\
&\\
\multicolumn{6}{c}{$E_B, \;2p_{-1}$ state}\\
0.1 &10&0.20084567 & 0.2008456\\
1  &16& 0.4565971  & 0.456597\\
2  &19& 0.599613 & 0.599613\\
20 &25&1.46551 & 1.4655\\
200&33& 3.3473 & 3.3471\\
300&35& 3.83485 & 3.8346\\
1000&38& 5.640 & 5.63842\\
\end{tabular}
\caption{Comparison of binding energies, $E_B = B(|m|+1)/2 - E$, of
the lowest energy states of
the $m$ = 0 and $-$1 manifolds obtained from the present $1/\kappa$
expansion, variational calculations
of R\"osner {\em et al.},\protect\cite{rosner}
and energy bounds of Handy {\em et al.}\protect\cite{handy}
Also given are estimates of the maximum perturbation
order for which double precision arithmetic suppresses roundoff
error sufficiently to yield the final coefficient with at least
one significant figure.}

\end{table}

\clearpage
\begin{center} {\large \bf Figures} \end{center}

Fig. 1. $\A_{jln}$ matrices arising from the first term in
Eq.(\protect\ref{eq:H_a}), for the first few orders $p$.
Superscripts indicate which recursion relation, if any, of
Eq.(\protect\ref{recurs}) may be used to compute that matrix, with
preference given to Eq.(\protect\ref{recur1}).

Fig. 2. Algorithm for applying recursion relations, using order $p=8$ as
an example.  Displayed is the state of the matrices $\A_{jln}$ after
each of the following steps: (a) status after order $p-1$; (b) apply
relation Eq.(\protect\ref{recur1}); (c) apply relation
Eq.(\protect\ref{recur1}); (d) compute three new elements.  Open and
filled circles denote matrices from order $p-1$ and $p$, respectively.

Fig. 3. Nonzero elements of the first few wavefunction expansion
matrices $\a_p$ for the ground state case.

Fig. 4. Layout of an $N \times N$ matrix on $M$ processors.

\def \X{\mbox{X}}
\clearpage
\begin{figure}
\begin{center}
\renewcommand{\arraystretch}{0.7}
\[
\begin{array}{lrrrrrr}
p=1 \hspace{0.5in} &\A_{310} \\
&\A_{300} \vspace{0.4in} \\
p=2 && \A_{420} \\
&\A_{311} & ^a\A_{410} \\
&\A_{301} & ^a\A_{400} \vspace{0.4in} \\
p=3 && \A_{421} & ^a\A_{520} \\
&\A_{312} & ^a\A_{411} & ^a\A_{510} \\
&\A_{302} & ^a\A_{401} & ^a\A_{500} \vspace{0.4in} \\
p=4 && & & ^b\A_{630} \\
&& \A_{422} & ^a\A_{521} & ^a\A_{620} \\
&\A_{313} & ^a\A_{412} & ^a\A_{511} & ^a\A_{610} \\
&\A_{303} & ^a\A_{402} & ^a\A_{501} & ^a\A_{600} \vspace{0.4in} \\
p=5 && & & ^b\A_{631} & ^a\A_{730} \\
&& \A_{423} & ^a\A_{522} & ^a\A_{621} & ^a\A_{720} \\
&\A_{314} & ^a\A_{413} & ^a\A_{512} & ^a\A_{611} & ^a\A_{710} \\
&\A_{304} & ^a\A_{403} & ^a\A_{502} & ^a\A_{601} & ^a\A_{700} \\
\vspace{0.4in} \\
p=6 && & & & & ^b\A_{840} \\
&& & & ^b\A_{632} & ^a\A_{731} & ^a\A_{830} \\
&& \A_{424} & ^a\A_{523} & ^a\A_{622} & ^a\A_{721}
&^a\A_{820} \\
&\A_{315} & ^a\A_{414} & ^a\A_{513} & ^a\A_{612} & ^a\A_{711}
&^a\A_{810} \\
&\A_{305} & ^a\A_{404} & ^a\A_{503} & ^a\A_{602} & ^a\A_{701}
&^a\A_{800} \\
\end{array}
\]
\end{center}
\vspace{0.5in}
\caption{}
\label{fig:matrices}
\renewcommand{\arraystretch}{1}
\end{figure}

\clearpage
\begin{figure}
\begin{center}
\begin{picture}(280,500)
\put(0,440){\LARGE a}
\put(0,340){\LARGE b}
\put(0,240){\LARGE c}
\put(0,140){\LARGE d}
\put(70,110){\vector(1,0){190}}
\put(70,210){\vector(1,0){190}}
\put(70,310){\vector(1,0){190}}
\put(70,410){\vector(1,0){190}}
\put(70,110){\vector(0,1){80}}
\put(70,210){\vector(0,1){80}}
\put(70,310){\vector(0,1){80}}
\put(70,410){\vector(0,1){80}}
\put(250,100){$j$}
\put(250,200){$j$}
\put(250,300){$j$}
\put(250,400){$j$}
\put(50,180){$l$}
\put(50,280){$l$}
\put(50,380){$l$}
\put(50,480){$l$}
\put(80,100){3}
\put(100,100){4}
\put(120,100){5}
\put(140,100){6}
\put(160,100){7}
\put(180,100){8}
\put(200,100){9}
\put(220,100){10}
\put(60,115){0}
\put(60,135){2}
\put(60,155){4}
\put(60,175){6}
\multiput(83,408)(20,0){8}{\line(0,1){4}}
\multiput(68,419)(0,10){7}{\line(1,0){4}}
\multiput(83,419)(20,0){7}{\circle{5}}
\multiput(83,429)(20,0){7}{\circle{5}}
\multiput(103,439)(20,0){6}{\circle{5}}
\multiput(143,449)(20,0){4}{\circle{5}}
\multiput(183,459)(20,0){2}{\circle{5}}

\multiput(83,308)(20,0){8}{\line(0,1){4}}
\multiput(68,319)(0,10){7}{\line(1,0){4}}
\multiput(103,319)(20,0){7}{\circle*{5}}
\multiput(103,329)(20,0){7}{\circle*{5}}
\multiput(123,339)(20,0){6}{\circle*{5}}
\multiput(163,349)(20,0){4}{\circle*{5}}
\multiput(203,359)(20,0){2}{\circle*{5}}
\put(83,319){\circle{5}}
\put(83,329){\circle{5}}
\put(103,339){\circle{5}}
\put(143,349){\circle{5}}
\put(183,359){\circle{5}}

\multiput(83,208)(20,0){8}{\line(0,1){4}}
\multiput(68,219)(0,10){7}{\line(1,0){4}}
\multiput(103,219)(20,0){7}{\circle*{5}}
\multiput(103,229)(20,0){7}{\circle*{5}}
\multiput(123,239)(20,0){6}{\circle*{5}}
\multiput(143,249)(20,0){5}{\circle*{5}}
\multiput(183,259)(20,0){3}{\circle*{5}}
\put(223,269){\circle*{5}}
\put(83,219){\circle{5}}
\put(83,229){\circle{5}}
\put(103,239){\circle{5}}

\multiput(83,108)(20,0){8}{\line(0,1){4}}
\multiput(68,119)(0,10){7}{\line(1,0){4}}
\multiput(83,119)(20,0){8}{\circle*{5}}
\multiput(83,129)(20,0){8}{\circle*{5}}
\multiput(103,139)(20,0){7}{\circle*{5}}
\multiput(143,149)(20,0){5}{\circle*{5}}
\multiput(183,159)(20,0){3}{\circle*{5}}
\put(223,169){\circle*{5}}
\end{picture}
\end{center}
\caption{}
\label{fig:algorithm}
\end{figure}

\clearpage
\begin{figure}
\renewcommand{\arraystretch}{0.7}
{\small
\tighten
\begin{center}
{\large $\a_0$} \hspace{0.5in}
$\left( \begin{array}{ccccccccc}
\X&0&0&0&0&0&0&0 \\
0&0&0&0&0&0&0&0 \\
0&0&0&0&0&0&0&0 \\
0&0&0&0&0&0&0&0 \\
0&0&0&0&0&0&0&0 \\
0&0&0&0&0&0&0&0&\cdots \\
0&0&0&0&0&0&0&0 \\
0&0&0&0&0&0&0&0 \\
0&0&0&0&0&0&0&0 \\
0&0&0&0&0&0&0&0 \\
 &&&&\vdots&&&&\ddots
\end{array} \right)$
\vspace{0.2in}

{\large $\a_1$} \hspace{0.5in}
$\left( \begin{array}{ccccccccc}
 0&0&0&0&0&0&0&0 \\
\X&0&\X&0&0&0&0&0 \\
0&0&0&0&0&0&0&0 \\
\X&0&0&0&0&0&0&0 \\
0&0&0&0&0&0&0&0 \\
0&0&0&0&0&0&0&0&\cdots \\
0&0&0&0&0&0&0&0 \\
0&0&0&0&0&0&0&0 \\
0&0&0&0&0&0&0&0 \\
0&0&0&0&0&0&0&0 \\
 &&&&\vdots&&&&\ddots
\end{array} \right)$
\vspace{0.2in}

{\large $\a_2$} \hspace{0.5in}
$\left( \begin{array}{ccccccccc}
 0&0&\X&0&\X&0&0&0 \\
 0&0&0&0&0&0&0&0 \\
\X&0&\X&0&\X&0&0&0 \\
0&0&0&0&0&0&0&0 \\
\X&0&\X&0&0&0&0&0 \\
0&0&0&0&0&0&0&0&\cdots \\
\X&0&0&0&0&0&0&0 \\
0&0&0&0&0&0&0&0 \\
0&0&0&0&0&0&0&0 \\
0&0&0&0&0&0&0&0 \\
 &&&&\vdots&&&&\ddots
\end{array} \right)$
\vspace{0.2in}

{\large $\a_3$} \hspace{0.5in}
$\left( \begin{array}{ccccccccc}
0&0&0&0&0&0&0&0 \\
\X&0&\X&0&\X&0&\X&0 \\
 0&0&0&0&0&0&0&0 \\
\X&0&\X&0&\X&0&\X&0 \\
0&0&0&0&0&0&0&0 \\
\X&0&\X&0&\X&0&0&0&\cdots \\
0&0&0&0&0&0&0&0 \\
\X&0&\X&0&0&0&0&0 \\
0&0&0&0&0&0&0&0 \\
\X&0&0&0&0&0&0&0 \\
 &&&&\vdots&&&&\ddots
\end{array} \right)$
\end{center}
}
\vspace{0.5in}
\caption{}
\label{fig:sparse}
\renewcommand{\arraystretch}{1}
\end{figure}

\clearpage
\begin{figure}
\begin{center}
\begin{picture}(300,300)
\multiput(50,50)(0,100){3}{\line(1,0){200}}
\multiput(50,50)(50,0){5}{\line(0,1){200}}
\put(20,150){\LARGE $N$}
\put(150,260){\LARGE $N$}
\put(70,215){\Large $n$}
\put(83,220){\vector(1,0){17}}
\put(67,220){\vector(-1,0){17}}
\put(121,190){\Large $\frac{N^2}{nM}$}
\put(125,210){\vector(0,1){40}}
\put(125,180){\vector(0,-1){30}}
\end{picture}
\end{center}
\vspace{0.4in}
\caption{}
\label{fig:layout}
\renewcommand{\arraystretch}{1}
\end{figure}


\begin{references}
\bibitem{note} No herculean programming effort was used to obtain
  this performance.  The entire code was written in standard CM
  Fortran.  It is possible that the use of utility library routines,
  scientific software library routines, or lower-level code might
  further improve this performance, but this has not been investigated.
\end{references}
\end{document}